\documentclass[twocolumn]{openjournal}
\usepackage{graphicx}	
\usepackage{amsmath}	

\begin{document}

\title{Type Ia supernovae observations do not show time dilation}
\shorttitle{No time dilation in type Ia supernovae}

\author{David F. Crawford}
\affil{Astronomical Society of Australia\\
Retired from School of Physics, University of Sydney.\\
44 Market St, Naremburn, 2065, NSW, Australia}
\email{davdcraw@gmail.com}

\begin{abstract}

The standard  analysis for type Ia supernovae uses a set of templates  to overcome the intrinsic variation of the supernova light curves with wavelength.
This paper shows that  standard templates  contain an anomaly in that  the width of the template light curve is proportional to the emitted wavelength.  Furthermore this anomaly is exactly what would be produced if epoch differences were not subject to  time dilation and yet  time dilation corrections were applied. It is the specific nature of this anomaly that is evidence for a static universe. The lack of time dilation is verified by direct analysis of the original supernovae data.
\end{abstract}


\maketitle

\section{Introduction}

It has been observed that type Ia supernovae are transient phenomena that take about ten days to reach a peak brightness and then the brightness  decreases at a slower rate. Type Ia supernovae (for brevity SNe) are also known for their remarkably constant absolute peak magnitudes and this property  makes them excellent cosmological probes.

The standard cosmological model of an expanding universe requires that the widths of the light curves must increase with redshift due to time dilation. The observed Hubble redshift, $z$,  is defined as the ratio of the observed wavelength to the emitted wavelength minus one. In an expansion model the ratio of any observed time  period to the emitted time period is identical to the ratio of the wavelengths, namely $(1+z)$. This is true for any time interval and is the time dilation. Any challenge to the standard model such as a static model must show that observations of SNe light-curve widths do not have time dilation even though the observed wavelengths  show a redshift.

The first strong evidence for time dilation in type Ia supernovae  was provided by \citet{Leibundgut96} with one supernova and \citet{Goldhaber96} with seven SNe. This was quickly followed by multiple SNe results from \citet{Goldhaber97,Perlmutter99,Goldhaber01}.  These papers record developments in both SNe observations and analysis, the results of which are asserted to provide strong evidence for  an expansion model chiefly because they show that the width of type Ia supernova light curves appears to increase with redshift in good agreement with an expanding model.

The results of this paper are based on the extensive analysis of  type Ia supernova observations provided by \citet{Betoule14} (hereafter B14).

There is an intrinsic variation of the shape of  light curves of SNe with emitted wavelength, which needs to be removed in order to measure the  peak luminosity and width for each supernova.  This removal is done  by comparing  the observations of each supernova to  a reference light curve obtained for each filter from a set of templates.

However  the set of templates used by B14 contains an anomaly in that the width of the reference light curve derived from these  templates is proportional to the rest-frame wavelength. It is argued that this relationship is an anomaly because it is completely unexpected in the standard cosmological model.   Importantly, this anomaly is exactly what would be produced if the epoch differences were not subject to  time dilation and yet time dilation corrections are applied.  It is the specific nature of this anomaly that is evidence for a static universe.

A reanalysis of the original SNe data  shows that the SNe data is fully consistent  with no time dilation. The implication is that the universe is static.

\section{The SNe data set}

Recently B14  have provided an update of the \citet{Conley11} analysis with better optical calibrations and more SNe. This JLA (Joint Light-curve  Analysis) list sample has 720 SNe  from the Supernova Legacy Survey (SNLS), nearby SNe (lowZ), the Sloan Digital Sky Survey (SDSS) \citep{Holtzman08,Kessler09} and those revealed by the Hubble Sky Telescope (HST) \citep{Riess07}.

\section{The B14 calibration method}
The definition of the redshift, $z$, is that  $(1+z=\lambda_0\lambda$, where $\lambda_0$ is the observed wavelength and  $\lambda$ is the rest-frame wavelength. The redshift is usually determined by the identification of known spectral lines in the observed spectrum. In an expanding model there is a cosmological time dilation in that any  observed time interval is $(1+z)$ times the emitted time interval. Clearly this time dilation is applicable to the light curves of supernovae and has a direct effect on the widths of their observed light curves.

The B14 calibration method \citep{Guy07} uses the SALT2 templates which provide the expected flux density of the supernova light curve as a function of both the rest-frame wavelength and the  difference between the observed epoch and the epoch of maximum response. The light-curve template file, Salt2\_template\_0.dat, provides the response for 20 days prior to the maximum and 50 days after the maximum for rest-frame wavelengths from 200 nm to 920 nm in steps of 0.5 nm.  The template file and  filter files for the JLA analysis were taken from the SNANA \citep{Kessler09} website. The prime purpose of this set of templates is to remove the effects of intrinsic variations that are common to all SNe. Assuming that the SNe are identical at all redshifts the beauty of the SALT2 method is that it uses the average of the light curves from all the SNe in order to compute the set of templates.

Let the cosmological time dilation be $f(1+z)$ then in the standard application the epoch differences in observations of a supernova are multiplied by $f(1+z)/(1+z)$. Provided the set of templates accurately reproduce (within measurement uncertainties) the expected light curve then  both the observed light curve  and the template light curve have been scaled by the same function. In effect the measured width is the ratio of the width of the observed light curve to the width of the template light curve.  Hence the measured  width of the light curve is independent of the value of this function. In other words the analysis method cannot distinguish between intrinsic widths, time dilation widths and time dilation corrections and all their effects are eliminated.  Thus the measured width is a valid estimate of the  width of the light curve for each supernova and its value is independent of the actual or assumed cosmology. For the standard analysis this width is called the stretch factor.

If $w_i(\lambda)$ is the intrinsic width of all SNe at the rest-frame wavelength $\lambda$ then the expected width of the light curve at observed wavelength $\lambda_0$ is
\begin {eqnarray}
\label{e1}
w(\lambda) & = &w_i(\lambda)f(1+z)/(1+z) \nonumber \\
w(\lambda)   & = & w_i(\lambda)f(\lambda_0/\lambda)(\lambda/\lambda_0)
\end {eqnarray}
where in the second line the definition of redshift has been used to replace the redshift by the wavelength ratio. For a given wavelength $\lambda_0$ the right hand side is a only a function of $\lambda$ which means that the set of templates can contain the time dilation function $f(\lambda)$ as well as the standard time dilation correction. Thus defining $w*$ to be the intrinsic width at a reference wavelength $\lambda*$  the relative width as a function of rest-frame wavelength is
\begin{equation}
\label{e2}
w/w*= f(\lambda^*/\lambda)(\lambda/\lambda^*)
\end{equation}

If the universe is static there is no time dilation (i.e. $f(\lambda)=1$) and this equation shows that for a static universe the  width of the templates produced by the standard analysis is essentially proportional to the  rest-frame wavelength (i.e $w\approx(w^*/\lambda^*)\lambda$). If the universe is expanding and the cosmological time dilation is cancelled by the time dilation correction then the light curves in the set of templates contains only  intrinsic effects. Hence the expected width distribution will be essentially constant. Since the measured widths of observed SNe are independent of cosmology the only cosmological information that can be derived from SNe widths in the B14 analysis is hidden in the set of templates that they used.

Thus the next step is to measure the light curve widths in the file Salt2\_template\_0.dat and investigate their relationship to the rest-frame wavelength. For nearby SNe  the observed wavelength is almost the same as the rest-frame wavelength and therefore the templates will be poorly determined in the gaps in wavelength between the filters. These gaps can be avoided by  using only the best covered wavelengths.

\begin{figure}
\includegraphics[width=0.47\textwidth]{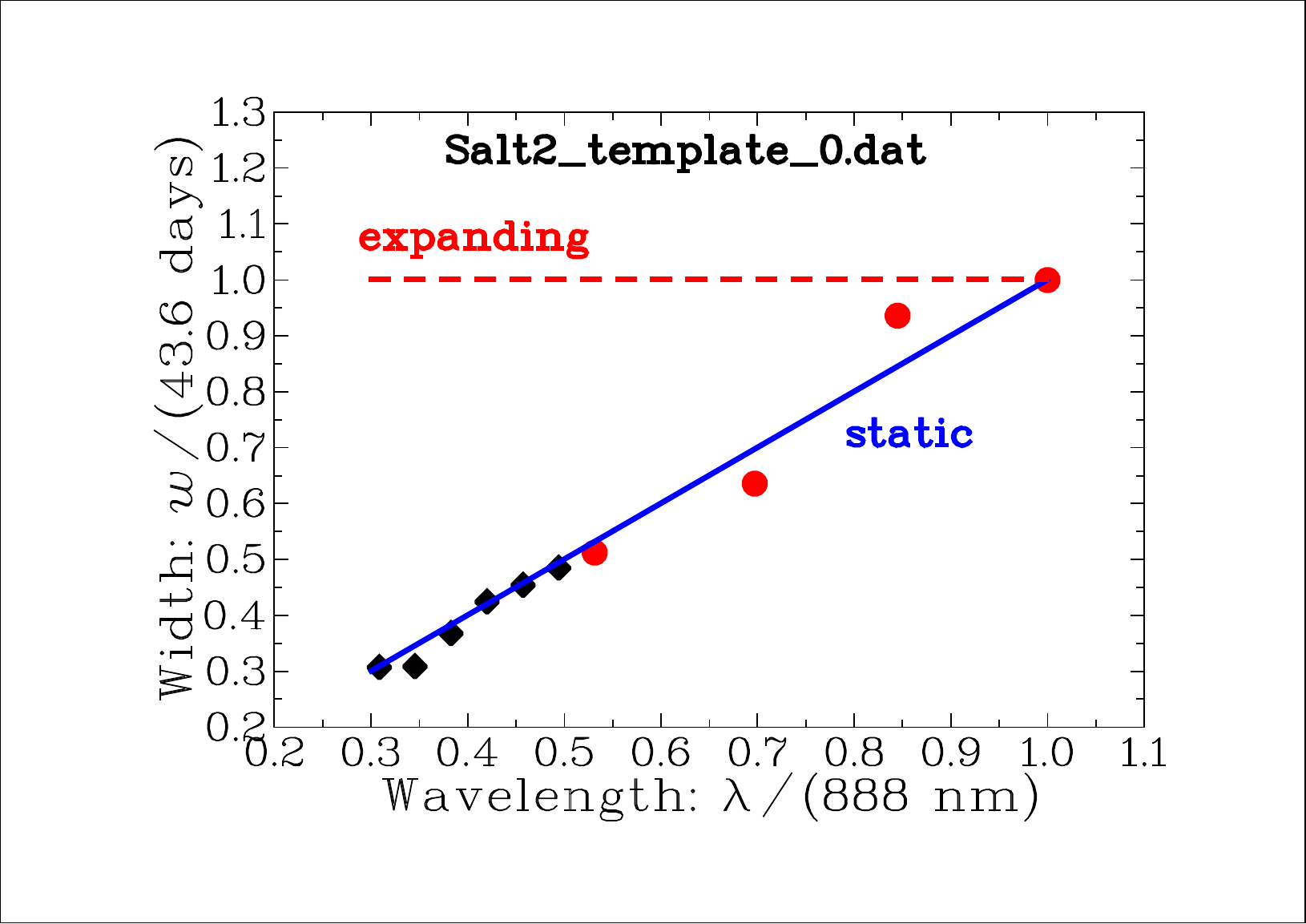}
\caption{ \label{f1} Plot of the relative  width ($w/43.6$ days) of  the SALT2 template SNe light curves as a function of the relative  rest-frame wavelength ($\lambda/888$ nm). The four widths (shown in red)  are for the $griz$ filters. The blue straight line is exactly what is expected if the supernovae did not have time dilation.  The red dashed line is what is expected if  the SNe have the expansion  time dilation of $(1+z)$}
\end{figure}

Light curves were determined from the SALT2 templates by using the filter gain curves for the $griz$ filters at zero redshift that is at 473 ($g$), 618 ($r$), 750 ($i$), and 888 ($z$) nm. Widths were measured for another six positions at the rest-frame wavelengths 274, 307, 340, 373, 406, 439 nm. In each  case the width was the epoch difference in days between the two half peak luminosity values of the light curve. Fig~\ref{f1} shows the results for the relative width as a function of the relative rest-frame wavelength $\lambda/\lambda^*)$.  For clarity the four filter positions have red symbols. The reference point is the central wavelength of the $z$ filter so that $\lambda^*=888$ nm and $w^*= 43.6$ days.  The solid blue is line is what is expected for a static universe. It is  exactly what would be produced if the supernovae  epoch differences did not have time dilation and yet standard time dilation corrections were applied. The red dashed line is what is expected if the universe is expanding. Clearly the light curve widths for the set of SALT2 templates favour the static universe.

The regression equation for the relative widths as a function of relative wavelength is
\begin{equation}
\label{e3}
w/w^*= (-0.023\pm0.025) + (1.036\pm0.044)(\lambda/\lambda^*)
\end{equation}
with a correlation coefficient of 0.992. Combining this result with  equation~\ref{e2} gives $f(\lambda^*/\lambda)=1.036\pm0.044 $) which favours a static universe.

If the anomaly is intrinsic then it should be observed in the relative widths between different filter observations for individual supernovae. In the SNANA data there are 45 SNe that have good observations in four or more filters and which have observations prior to the maximum. For each of these SNe a  least squares fit was done for  flux densities observed for each filter against a common template to get a width for each filter. Note that the same template was used for all filters. Then using the central wavelengths for each filter a simple regression  over all filters provides the slope of the width as a function of  rest-frame wavelength. For all the SNe, weighted average slope was $-0.095\pm0.116$ . If the anomaly was intrinsic the slope should be near one. Hence the anomaly can not be due to an intrinsic property of the SNe.

\section{Re-analysis of original observations}
The original observations were then analysed  to see if they are consistent with a static universe.  Where necessary the time dilation  correction is removed from the reference light curve (using the SALT2 templates) for each filter by numerical interpolation.  This preserves the intrinsic wavelength dependence of the light curves but eliminates the unwarranted time dilation corrections in the template. A least squares fit width estimation program was run for all SNe that were readily available on the SNANA site and had estimates for the epoch of peak luminosity. The major difference from the B14 sample is that 228  SNe from the SDSS survey were not easily accessible and since they were all in the lower redshift region their omission makes little difference to the results. For each of the 519 supernovae there was a weighted least squares fit for the peak flux density from the flux densities for each filter and then
a least squares fit to a common width estimate for all filters.

The slopes of the light curve width  as a function of $(1+z)$  were done for two cases. The first is a repeat of the standard  analysis where time dilation corrections are consistently applied. The slope of the relative width as a function of $(1+z)$ is $-0.093 \pm 0.049$ in  agreement  with B14.

The second case was done without time dilation corrections being applied to the observed data and with the SALT2 templates being corrected (numerically) for the anomalous  time dilation correction  that had been applied.  Here the slope is  $-0.096 \pm 0.050$ that is consistent with zero. Note that if the SNe have a cosmological time dilation equal to $(1+z)$ this slope would be close to unity. Thus the reanalysis of the original observations favour a static universe.

\section{Conclusions}

The normal Hubble variation of observed wavelength with redshift is well established and in an expanding universe time dilation should show the identical dependence. Observations of type Ia supernovae are  one of the few observations that can directly show time dilation. However this paper shows that observations of type Ia supernovae do not show the effects of time dilation. The implications of measurements of peak magnitudes in a static universe are considered in a later paper.  The major conclusion of this paper is that  there is no time dilation in the widths of type Ia supernovae light curves and consequently the universe is static. A second result is that the measured width is a valid estimate of the  width of the light curve for each supernova and its value is independent of the actual or assumed cosmology.

\section{Acknowledgements}
This research has made use of NASA's Astrophysics Data System Bibliographic Services. The calculations have been done using Ubuntu Linux and the graphics have been done using the DISLIN plotting library provided by the Max-Plank-Institute in Lindau.



\end{document}